\documentclass[12pt]{article}

\advance \textheight by \topskip
\def\N{\mathbb N}
\def\Z{\mathbb Z}

\def\cR{\mathcal R}
\def\I{\hbox{{1\hskip -5.8pt 1}\hskip -3.35pt I}}

\def\ket #1{|{#1}\rangle}

\def\mop #1{\mathop{\sf #1}\nolimits}

\usepackage{amsmath,amssymb,euscript,graphicx}

\begin{document}
\title{
{\bf Matrix equations and trilinear commutation relations}}
\author{{\sf Sergey Klishevich}\thanks{E-mail:
klishevich@ihep.ru}\\
{\small {\it Institute for High Energy Physics, Protvino,
Russia}}}
\date{}
\maketitle

\vskip-1.0cm

\begin{abstract}
  In this paper we discuss a general algebraic approach to
  treating static equations of matrix models with a
  mass-like term. In this approach the equations of motions
  are considered as consequence of parafermi-like trilinear
  commutation relations. In this context we consider several
  solutions, including construction of noncommutative
  spheres. The equivalence of fuzzy spheres and parafermions
  is underlined.
\end{abstract}
\newpage

\section{Introduction}

The BFSS and IKKT matrix models \cite{BFSS,Kawai} are
widely used in the context of string theories since it is
believed that matrix models nonperturbatively describe
collective degrees of freedom --- branes (see also
Ref.~\cite{soch}).
It was conjectured that superstring theories with brane
degrees of freedom correspond to perturbative regimes of
M-theory, which pretends to be {\it theory of everything}.
In the light-cone frame it was conjectured to be described
be an Yang-Mills type matrix mechanics (BFSS matrix model)
\cite{BFSS}.
The IKKT matrix model was offered as an effective theory for
the large-N reduced model of super Yang-Mills theory while
solutions of the model were interpreted as infinitely long
static D-string configurations \cite{Kawai,Li96}.

Static equations of matrix models are very important since
their solutions are used as an background (an vacuum
configuration) when quantizing system with specific
configuration (e.g. see Ref. \cite{Kimura01,Kimura01a}).
Therefore for applications it is important to
have so many such solutions as possible. In this paper we
offer a general algebraic approach to treating static
equations of motions of matrix models and discuss various
solutions.

The paper is organized as follows. In the next section we
offer an algebraic approach to treating static matrix
equations with a mass-like term which can be interpreted as
an interaction with Ricci tensor in the case of curved
space. In this approach the matrix equations of motions are
considered as constraints on parafermi-like trilinear
commutation relations. In the section \ref{linc} we discuss
several solutions corresponding to the linear case,
including construction of fuzzy spheres. Besides we have
demonstrated equivalence of parafermions and noncommutative
spheres. Brief discussion of results is presented in section
\ref{conc}.

\section{An algebraic approach to static matrix equations}

Let us consider the matrix model given by the action
\begin{equation}\label{S2}
 S =\int dt\mop{Tr}
 \left(\dot X_\mu\dot X^\mu-
 \frac 12\left[X_\mu,\,X_\nu\right]
 \left[X^\mu,\,X^\nu\right]
 - R_{\mu\nu}X^\mu X^\nu\right).
\end{equation}

Here coordinates $X^\mu$ are $N\times N$ Hermitian matrices,
all the components $R_{\mu\nu}$ are numbers while Greek
indices run form $1$ to $p$. Repeated upper and lower
indices are implicitly summed over while for raising and
lowering of indices the metric $g_{\mu\nu}$ is used. The
last term can be interpreted as a generally non-diagonal
mass
term or as a term representing the interaction of the vector
field $X_\mu$ with an external symmetric field $R_{\mu\nu}$
(e.g. Ricci tensor in a space with nontrivial curvature).
Advantage of such matrix models is that they admit stable
vacuum solutions, which can be interpreted as compact branes
(e.g. spherical branes). In the matrix model picture, $N$
represents the number of the quantums on the backgrounds (or
the number of D-instantons or D-particles). Without the mass
term the model \eqref{S2} is the bosonic part of the BSFF
matrix model \cite{BFSS} however in principle the tensor
$R_{\mu\nu}$ can depend on $N$ and vanish in large $N$
limit. Later on we will discuss classical solutions only,
therefore, we discard the fermionic degrees of freedom.

The action (\ref{S2}) leads to the following equations:
\begin{equation}\label{teq}
[X^\nu,\,[X_\nu,\,X_\mu]] - \ddot X_{\mu} = R_{\mu\nu}X^\nu.
\end{equation}

The equations of motions for static solutions have the
algebraic form of trilinear commutation relations:
\begin{equation}\label{YM}
 [X^\nu,\,[X_\nu,\,X_\mu]] = R_{\mu\nu}X^\nu.
\end{equation}
Solutions to these equations should correspond to
even-dimensional D-p-branes.

If we treat the metric as pseudo-Euclidean then the
relations \eqref{YM} correspond to equations of motions of
IKKT matrix model with a mass like term.
The IKKT matrix model with a mass term was
considered in Ref.~\cite{Kimura01}.

It is worth noting that the nontrivial right-hand side in
\eqref{YM} make the algebra generated by coordinates $X^\mu$
to be different from the homogeneous case \cite{connes}.

From the algebraic point of view we can treat the matrix
equations of motion \eqref{YM} as constraints on a priori
nonlinear algebra defined by the trilinear relations
\begin{align}\label{gA}
 [X_\rho,\,[X_\mu,\,X_\nu]]&=R_{\rho\mu\nu}(X,\mathcal A),
\end{align}
where $\mathcal A$ denotes a set of operators independent of
the coordinates $X_\mu$. The r.h.s. has the obvious index
symmetries, which follow from the properties of the double
commutator on the l.h.s. Such trilinear relations can be
thought as a generalization of the trilinear commutation
relations for parafermi systems \cite{kam}.

To link the commutation relations \eqref{gA} with matrix
equations of motions \eqref{YM} tensor operator
$R_{\nu\rho\mu}(X,\mathcal A)$ has to obey the condition
\begin{align}\label{constr}
  g^{\nu\rho}R_{\nu\rho\mu}(X,\mathcal A)\sim
  R_{\mu\nu}X^\nu
\end{align}
So, the whole algebra is generated by the coordinates
$X_\mu$ and the set $\mathcal A$ governed by the trilinear
relations \eqref{gA} and the constraints \eqref{constr}.
Since we assume that the operators from the set $\mathcal A$
do not appear in the equations \eqref{YM} they can be
treated as objects representing ``topological'' degrees of
freedom (e.g. charges corresponding to D-branes) of the
matrix model.

It is necessary to note that the offered approach is general
because any solution to the equations \eqref{YM} obeys the
relations \eqref{gA} and \eqref{constr} for some tensor
operator $R_{\nu\rho\mu}(X,\mathcal A)$ since the equations
\eqref{gA} can be taken as its definition.

The relations \eqref{gA} are too general and we restrict
our self to the case when the set $\mathcal A$ is trivial
(proportional to identity or empty). When it is natural to
consider $R_{\nu\mu\rho}(X)$ as a regular function in
$X_\mu$, i.e. it can be represented as a series
$$
 R_{\rho\mu\nu}(X)=R_{\rho\mu\nu}^{(0)} +
 R^\lambda{}_{\rho\mu\nu}X_\lambda + \ldots,
$$
where in this case the constant tensor
$R_{\nu\mu\rho}^{(0)}$ is related to central charges of an
algebra. The nonlinear dependence on the coordinates $X_\mu$
in general leads to nonlinear algebras but in this paper we
discuss the linear case only.

\section{Linear case}
\label{linc}

Later on we will discuss the partial case, when the
r.h.s. \eqref{gA} is linear in the coordinates $X_\mu$:
\begin{align}\label{lin}
  [X_\rho,\,[X_\mu,\,X_\nu]]&=
  R^\lambda{}_{\rho\mu\nu}X_\lambda,
\end{align}
where $R^\lambda{}_{\nu\mu\rho}$ commutes with the all
coordinates and will be treated as a constant tensor
related to structure constants of the Lie algebra generated
by the coordinates $X_\mu$. Besides, here we suppose that
the coordinates generate all operators in the space.

The tensor $R^\lambda{}_{\nu\mu\rho}$ has the usual index
symmetry of a Riemann tensor
$$
 R^\lambda{}_{\rho(\mu\nu)}=0,\hskip 2cm
 R^\lambda{}_{\rho\mu\nu}+R^\lambda{}_{\mu\nu\rho}
 +R^\lambda{}_{\nu\rho\mu}=0.
$$
We will see that in some cases this tensor indeed is a
Riemann tensor of a symmetric space with the corresponding
algebra defined by the trilinear commutations relations
\eqref{lin}. Therefore, we will adopt this terminology for
it. It is interesting to note that according to this
terminology the massless solutions correspond to Ricci flat
spaces.

Let us consider the following three cases:

\begin{enumerate}

\item
Let the coordinates $X_\mu$ form a basis of some simple
algebra $\mathfrak a$. Then the Riemann tensor has the form
$$
 R^\lambda{}_{\nu\mu\rho}=
 C^\lambda_{\nu\sigma}C^\sigma_{\mu\rho},
$$
where $C^\lambda_{\nu\sigma}$ are structure constants of the
simple algebra. This solution can be identified with a
D-brane. If one takes a semi-simple algebra, then such a
solution corresponds to a collection of D-branes.

\item
If the noncommutative coordinates $X_\mu$ form only a part
of basis of some simple algebra then this algebra has a
$\Z_2$-grading defined by the relation $\mop{gr}X_\mu=1$.
Therefore, such an algebra can be decomposed as
$\mathfrak p\oplus\mathfrak m$, where
$\mathfrak p=\mop{span}\{X_\mu\}$ and
$\mathfrak m=\mop{span}\{[X_\mu,\,X_\nu]\}$, with the
structural relations
\begin{align}\label{sr}
 [\mathfrak m,\,\mathfrak m]&\subset\mathfrak m,&
 [\mathfrak m,\,\mathfrak p]&\subset\mathfrak p,&
 [\mathfrak p,\,\mathfrak p]&\subset\mathfrak m.
\end{align}
The subspace $\mathfrak p$ can be identified with a tangent
space of the corresponding symmetric space. On the other
hand one can represent the commutators of the coordinates
in the form
$$
 [X_\mu,\,X_\nu]=\frac 12R^\lambda{}_{\rho\nu\mu}
 M_\lambda{}^\rho,
$$
where
$M_\lambda{}^\rho\in\mathfrak{so}(p)\supseteq\mathfrak m$,
therefore, the noncommutative coordinates can be associated
with the covariant derivatives on the symmetric space.

\item
In general, it is not forbidden for the algebra generated by
the operators $X_\mu$ subjected to \eqref{lin} to be
infinite-dimensional. Later we will briefly discuss an
explicit example to this case.

\end{enumerate}

Some comments are in order. The coordinates $X_\mu$ can be
represented by finite $N\times N$ matrices iff the algebras
in the first two cases are compact. Since a Lie group is a
symmetric space as well, hence, the first case can be
included into the second. Therefore, we will not discuss
this case here. In the cases corresponding to symmetric
spaces the metric should be proportional to the Ricci
tensor.

\subsection{Parafermions and noncommutative spheres}

Let us start with the case of a constant curvature space,
which is the most simple case among symmetric spaces. The
Riemann tensor has the form
\begin{equation}\label{ccs}
 R^\lambda{}_{\rho\mu\nu}=\frac{\cR}{p(p-1)}
  (\delta^\lambda_\mu g_{\nu\rho} -
   \delta^\lambda_\nu g_{\mu\rho}),
\end{equation}
where $\cR$ is a scalar curvature. We look for the
solution for the coordinates in the form of the rescaled
operators
$$
 X_\mu=\frac 12 \sqrt{\frac{\cR}{p(p-1)}}G_\mu,
$$
where Hermitian operators $G_\mu$ obey the commutations
relations
\begin{equation}\label{gpf}
 [G_\rho,\,[G_\mu,\,G_\nu]]=
  4(g_{\nu\rho}G_\mu - g_{\mu\rho}G_\nu).
\end{equation}
The operators $G_\mu$ generate the algebra
$\mathfrak{so}(p+1)$, $p\in\N$:
\begin{align}
 \left[G_\mu,\,G_\nu\right]&=4iM_{\mu\nu},&
 \notag\\\label{AdS}
 \left[G_\mu,\,M_{\nu\rho}\right]&=
 i\left(g_{\mu\nu}G_\rho - g_{\mu\rho}G_\nu\right),
 \\\notag
 \left[M_{\mu\nu},\,M_{\rho\sigma}\right]&=
 i\left(g_{\mu\rho}M_{\sigma\nu}
      - g_{\mu\sigma}M_{\rho\nu}
      + g_{\nu\sigma}M_{\rho\mu}
      - g_{\nu\rho}M_{\sigma\mu}\right).
\end{align}

\subparagraph{Even dimensional case.}
Let us consider the case of even $p$. Then the Hermitian
operators $G_{\mu}$ can be represented in the following
form:
\begin{align*}
 && G_{2k-1}&=a^\dag_k+a_k,&
 G_{2k}&=i(a^\dag_k-a_k),&&
\end{align*}
where the mutually conjugate operators $a_k$, $a^\dag_k$
obey the basic parafermi commutation relations \cite{kam}
\begin{align}\label{pf}
 [a_k,\,[a^\dag_l,\,a_m]]&=2g_{kl}a_m, &
 [a_k,\,[a^\dag_l,\,a^\dag_m]]&=
 2g_{kl}a^\dag_m-2g_{km}a^\dag_l, &
 [a_k,\,[a_l,\,a_m]]&=0.
\end{align}
So, this case corresponds to a parafermi system.
For parafermi systems the most important representation is
the Fock representation defined by the relations:
\begin{equation}\label{pFock}
 a_k\ket 0=0,\hskip 2cm
 a_ka^\dag_l\ket 0=n\delta_{kl}\ket 0,
 \qquad n\in\N.
\end{equation}
This representation is labeled by the natural number $n$
called the order of the parafermi quantization. This
representation can also be given in terms of the Green
ansatz \cite{kam}:
\begin{align}\label{Gr0}
 G_\mu&=\sum_{\alpha=1}^n G_\mu^{(\alpha)},
\end{align}
where the summands obey the relations
\begin{align}\label{Gr1}
 \left\{G_\mu^{(\alpha)},\,G_\nu^{(\alpha)}\right\}&=
 2g_{\mu\nu}\cdot\I,&
 \left[G_\mu^{(\alpha)},\,G_\nu^{(\beta)}\right]&=0,
 \qquad\alpha\ne\beta.
\end{align}
The natural matrix representation of the Green ansatz is
given on the vector space $V_a=V^{\otimes^n}$:
\begin{align}\label{GrM}
 G_\mu^{(1)}&=\Gamma_\mu\otimes\I\otimes\cdots\otimes\I,&
 G_\mu^{(2)}&=\I\otimes\Gamma_\mu\otimes\cdots\otimes\I,&
 \ldots&,&
 G_\mu^{(n)}&=\I\otimes\cdots\otimes\I\otimes\Gamma_\mu,
\end{align}
where $V$ is the irreducible representation space of the
Clifford algebra
\begin{equation}\label{cliff}
  \{\Gamma_\mu,\,\Gamma_\nu\}=2g_{\mu\nu}\cdot\I.
\end{equation}

The operators $G_\mu^{(\alpha)}$ can be represent as a
linear combination of the fermi creation-annihilation
operators:
\begin{align*}
 G^{(\alpha)}_{2k-1}&=
  a^{(\alpha)}_k{}^\dag+a_k^{(\alpha)},&
 G^{(\alpha)}_{2k}&=
  i(a^{(\alpha)}_k{}^\dag-a^{(\alpha)}_k).
\end{align*}
The vector space $V_a$ is the corresponding Fock space
generated by polynomials of the operators
$a^{(\alpha)}_k{}^\dag$ acting on the vacuum. The
symmetrized traceless vector space
$V_{st}=\mop{Sym}_tV^{\otimes^n}$ is a proper subspace of
$V_a$, which contains the vacuum of the Fock representations
on $V_a$ and $V_{st}$. Besides, this symmetrized space is
invariant with respect to the action of the operators
\eqref{Gr0}. The Fock representation of the parafermi system
\eqref{pf}, \eqref{pFock} belongs to the space $V_{st}$
since the corresponding vacuum state does. It is known
Ref.~\cite{Greenberg} that the irreducible Fock
representation of the parafermi relations is a proper
subspace of $V_a$, therefore, since the vacuum belong to
$V_{st}$ this space coincides with the irreducible
representation space of the parafermi system. This proofs
the following theorem.

{\it Theorem:}
The matrix representation \eqref{Gr0}-\eqref{cliff} on the
symmetrized traceless vector space
$\mop{Sym}_tV^{\otimes^n}$ is the irreducible Fock
representation of the parafermi commutation relations of the
order $n$.

{\bf Time dependent solution.}
Using the one-parametric unitary transformation of the Fock
representation
\begin{equation*}
 a_k\to a_ke^{-i\omega t},\hskip 2cm
 a_k^\dag\to a_k^\dag e^{i\omega t}
\end{equation*}
we can write down the time dependent solution of the matrix
equations \eqref{teq} in terms of the time dependent
matrices
\begin{equation}\label{time}
 G_\mu=\sum_{k=1}^{p/2}\left(
 B_\mu^k a_ke^{-i\omega t}
 + (B_\mu^k)^*a^\dag_ke^{i\omega t}\right),
\end{equation}
where
$$
  B_{2k-1}^l=\delta_{kl},\hskip 2cm
  B_{2k}^l=i\delta_{kl}.
$$
The coordinates can be represented in the form
\begin{equation}\label{pFX}
  X_\mu=\frac r{2\sqrt{p(p-1)}}\,G_\mu,
\end{equation}
where from the \eqref{gpf} in follows that the parameter $r$
is defined by the relation
$$
 r^2=\cR-p\omega^2.
$$
Since the parameter $r$ has to be positive real number
the scalar curvature and frequency are restricted:
$\cR>0$ and $\omega^2<\cR/p$.

Besides the harmonic time dependence we can consider the
coordinates with exponential dependence,
$\ddot X_{\mu}-\omega^2X_{\mu}=0$. In this case there is no
constraints on the parameter~$r$.

{\bf Large $N$ limit.} It is known that for the Fock
representations of the parafermi system the rescaled
operators $a_k/\sqrt{n}$ and $a^\dag_k/\sqrt{n}$ in the
limit $n\to\infty$ go to the Heisenberg algebra. Therefore,
in this limit the coordinates \eqref{pFX} have the
commutation relation
$[X_\mu,\,X_\nu]=\theta_{\mu\nu}\cdot\I$. In other words,
the noncommutative manifold defined by the coordinates
\eqref{pFX} in large $N$ limit goes to the fuzzy plane
$\mathbb R_\theta^p$ representing a noncommutative
D-p-brane. This solution is a BPS one \cite{BFSS}.

In the paper \cite{4sph} the representation
\eqref{Gr0}-\eqref{cliff} on the symmetrized vector space
was used to realize the noncommutative 4-sphere. Later
similar representations of higher even-dimensional fuzzy
spheres were realized on the symmetrized traceless vector
space $\mop{Sym}_tV^{\otimes^n}$ \cite{Ramgoolam, 0306250}.
Thus as we have seen the irreducible Fock representations of
parafermions are equivalent to noncommutative spheres.

\subsection{About solutions with a nontrivial Weyl tensor}

Above we discussed the cases related to the symmetric spaces
with a trivial Weyl tensor. However, the non-triviality of
the Weyl tensor is important for constructing ``massless''
solutions (with $R_{\mu\nu}=0$). Indeed, in the trilinear
relations \eqref{lin} one can change the Riemann tensor for
the Weyl one
\begin{align}\label{Weyl}
 [X_\rho,\,[X_\mu,\,X_\nu]]&=
  W^\lambda{}_{\rho\mu\nu}X_\lambda
\end{align}
while the metric is kept the same. Obviously, the new
trilinear commutation relations \eqref{Weyl} will provide
solutions of the necessary type.

Let us consider the case of the symmetric space
$SU(n)/SO(n)$. The basis of the algebra $\mathfrak{su}(n)$
can be represented by matrices $s_{ab}+im_{ab}$, where
$s^\dag_{ab}=s_{ab}$, $s_{ab}=s_{ba}$ and
$m_{ab}$ span a Hermitian basis of the algebra
$\mathfrak{so}(n)$. We will identify the operators $s_{ab}$
with the coordinates $X_\mu$, i.e. in this case $\mu$ is a
multi-index. The commutation relations between
$s_{ab}$ and $m_{ab}$ are
\begin{align*}
 [s_{ab},\,s_{cd}]&=-2i
 (\delta_{c(a}m_{b)d} + \delta_{d(a}m_{b)c}),&
 [s_{ab},\,m_{cd}]&= 2i
 (\delta_{c(a}s_{b)d} - \delta_{d(a}s_{b)c}).
\end{align*}
Therefore, the Riemann tensor in \eqref{lin} has the form
\begin{equation}\label{Rsu_n}
 R^{(lm)}{}_{(ef)(ab)(cd)}\sim 4\left(
     \delta^{\strut}_{(c|(a}\delta^{\strut}_{b)(e}
     \delta_{f)}^{(l}\delta_{|d)}^{m)}
     - \delta^{\strut}_{(a|(c}\delta^{\strut}_{d)(e}
     \delta_{f)}^{(l}\delta_{|b)}^{m)}\right).
\end{equation}
The Ricci tensor is proportional to the metric
$$
 R_{(ab)(cd)}\sim g_{(ab)(cd)}=\delta_{a(c}\delta_{d)b}
 - \frac 1n\delta_{ab}\delta_{cd}.
$$
Comparing the Riemann tensor \eqref{Rsu_n} with the Ricci
one it is possible to conclude that the corresponding Weyl
tensor is nontrivial, therefore nontrivial trilinear
commutation relations \eqref{Weyl} providing a massless
solution can be constructed.

\subparagraph{The case of $A$-statistics.} In this case the
basic commutation relations are given in the form
\cite{{A-stat}}
\begin{align}\label{A-stat}
  &[[a_k^\dag,\,a_l],\,\bar a_m] =
  \delta_{kl}a_m^\dag + \delta_{lm}a_k^\dag,\notag\\
  &[[a_k,\,a_l^\dag],\,a_m] =
  \delta_{kl}a_m + \delta_{lm}a_k,\\\notag
  &[a_k,\,a_l]=[a_k^\dag,\,a_l^\dag]=0,
\end{align}
where  $k,l,m=1,\ldots,M$.

In terms of the basis of the Hermitian operators
$X_\mu$:
$$
 X_k=a_k^\dag+a_k,\hskip 2cm
 X_{n+k}=i(a_k^\dag-a_k)
$$
the commutation relations \eqref{A-stat} can be represented
in the form \eqref{lin} with the Riemann tensor
\begin{equation}\label{AsR}
 R^\alpha{}_{\beta\mu\nu} =
   \delta_\mu^\alpha g_{\beta\nu}
 - \delta_\nu^\alpha g_{\beta\mu}
 + J_\mu^\alpha J_{\beta\nu}
 - J_\nu^\alpha J_{\beta\mu}
 + 2 J_\beta^\alpha J_{\mu\nu},
\end{equation}
where
\begin{align*}
 g_{\mu\nu}&=\begin{pmatrix}
              \I_{M\times M}&0\\0&\I_{M\times M}
             \end{pmatrix},&
 J_{\mu\nu}&=\begin{pmatrix}
              0&-\I_{M\times M}\\\I_{M\times M}&0
             \end{pmatrix}.
\end{align*}
The commutation relations \eqref{lin} with this Riemann
tensor coincide with those of covariant derivatives on the
complex projective plane $\mathbb{C}P^M$. The tensor
$J_{\mu\nu}$ is a complex structure tensor on this space. It
is easily to see that the associated Weyl tensor is
nontrivial in this case.

Consider a Fock representation for the A-statistics
\eqref{A-stat} given by polynomials in operators $a_k\dag$
over a ground state $\ket 0$ formally defined by the same
relations \eqref{pFock}. In this case the natural parameter
$n$ is called as an order of the $A$-statistics. This
representation is finite-dimensional and in the large $N$
limit ($n\to\infty$) this representation goes to that of
Heisenberg algebra (a noncommutative plane) like the
parafermionic representation does.

\subsection{The generalized Dolan-Grady relations}

The interrelationship between the equations (\ref{YM}),
one-mode parafermion and the Dolan-Grady relations was noted
in Ref. \cite{Dolan-Grady}. Here we discuss application of
the generalized Dolan-Grady relations \cite{ivanov}.

The real form of the generalized Dolan-Grady relations
can be represented as
\begin{align}\label{gDG}
 [G_\nu,\,[G_\mu,\,G_\nu]]&=\epsilon_\nu G_\mu,&
 [G_\rho,\,G_\sigma]&=0,&
 \epsilon_\nu&=\pm 1,
\end{align}
where $G_\mu$ are Hermitian operators, the indices run from
$1$ to $p$, there is no summation over the index $\nu$, the
indices $\mu$ and $\nu$ are supposed to acquire adjacent
values while the second identity is supposed for the case
when the indices $\rho$ and $\sigma$ are not adjacent. The
values $1$ and natural $p$ are considered as adjacent.

However, it is worth noting that if
one takes the metric in the form
$g_{\mu\nu}=\delta_{\mu\nu}$ when for the coefficients
$\epsilon_\nu$ obeying the relation
$$
 \epsilon_i+\epsilon_{i+2}=0,
$$
where we imply the periodic condition
$\epsilon_{i+p}=\epsilon_i$,
the generalized Dolan-Grady relations provide a
stationary solution to the usual BFSS model (massless
solution). Such a solution exists only for $p$ to be a
number divisible by $4$. So, this solution should correspond
to even D-p-branes as it is required.

The generalized Dolan-Grady relations define
$\mathfrak{sl}(p)$ Onsager algebra \cite{ivanov}.
Unlike the usual Onsager algebra the theory of
finite-dimensional representations in the general case of
$\mathfrak{sl}(p)$ Onsager algebra is not elaborated. Till
now the only known such representations can be defined in
terms of the operators
\begin{align*}
  g_{2k-1}&=\frac 12\sigma^x_k,&
  g_{2k}&=\frac 12\sigma^z_k\sigma^z_{k+1},
\end{align*}
where $k=1,2,\ldots,L$. The case $G_\mu=g_\mu$ corresponds
to the $\mathfrak{sl}(2L)$ Onsager algebra, while for
$2L=mp$ ($m\in\N$) the operators of the form
$$
 G_\mu=\sum_{s=0}^{m-1}g_{\mu+ps},
$$
where $\mu=1,\ldots,p$, generate the $\mathfrak{sl}(p)$
Onsager algebra \cite{ivanov}. The limit $L\to\infty$ (or
more specifically $m\to\infty$ while $p$ is fixed)
corresponds to the large $N$ limit.

One can consider more restrictive commutation relations then
the generalized Dolan-Grady relations \eqref{gDG} taking
the ``Riemann tensor'' of the form
$$
 R^\lambda{}_{\nu\mu\rho}\sim\epsilon_\nu\bigr(
 \delta^\lambda_\mu\delta_{\rho\nu}
 (\delta_{\mu,\nu+1}+\delta_{\mu,\nu-1})
 - \delta^\lambda_\rho\delta_{\mu\nu}
 (\delta_{\rho,\nu+1}+\delta_{\rho,\nu-1})
 \bigl).
$$
The corresponding ``Ricci tensor'' is
$R^\lambda{}_{\nu\lambda\rho}\sim
 \epsilon_\nu\delta_{\nu\rho}$, therefore, in the case of
the BFSS model it cannot be taken as a metric if
$\epsilon_\nu$ have different signs.

The relations \eqref{gDG} can also be used to construct
solutions of the IKKT model. In this case there is the same
restriction on $p$.

\section{Conclusion}
\label{conc}

In this paper we offered a general algebraic approach to
treating static equations of motions of Yang-Mills type
matrix models. We have considered several solutions,
including discussed in the literature construction of higher
dimensional noncommutative spheres. We underlined
equivalence of fuzzy spheres and parafermions and
demonstrated that this parafermionic solution can be
extended to include harmonic time dependence but with
restrictions on the scalar curvature and frequency.

Relying on the noted analogy between the equations of motion
of Yang-Mills type matrix models and the parafermi-like
commutation relations we hope that the numerous ideas and
methods elaborated for parastatistics (e.g. the Green
representation) will be useful in developments related to
the matrix equations.


\end{document}